\documentclass[epj]{svjour}
\usepackage{graphicx}

\def\mb#1{\setbox0=\hbox{$#1$}\kern-.025em\copy0\kern-\wd0
\kern-0.05em\copy0\kern-\wd0\kern-.025em\raise.0233em\box0}

\begin{document}
   \title{Lynden-Bell and Tsallis distributions for the HMF model}

 \author{P.H. Chavanis}

\institute{ Laboratoire de Physique Th\'eorique, Universit\'e Paul
Sabatier, 118 route de Narbonne 31062 Toulouse, France\\
\email{chavanis@irsamc.ups-tlse.fr}}

\titlerunning{}

   \date{To be included later }

   \abstract{Systems with long-range interactions can reach a Quasi
   Stationary State (QSS) as a result of a violent collisionless
   relaxation. If the system mixes well (ergodicity), the QSS can be
   predicted by the statistical theory of Lynden-Bell (1967) based on
   the Vlasov equation. When the initial condition takes only two
   values, the Lynden-Bell distribution is similar to the Fermi-Dirac
   statistics. Such distributions have recently been observed in
   direct numerical simulations of the HMF model (Antoniazzi et
   al. 2006). In this paper, we determine the caloric curve
   corresponding to the Lynden-Bell statistics in relation with the
   HMF model and analyze the dynamical and thermodynamical stability
   of spatially homogeneous solutions by using two general criteria
   previously introduced in the literature. We express the critical
   energy and the critical temperature as a function of a degeneracy
   parameter fixed by the initial condition. Below these critical
   values, the homogeneous Lynden-Bell distribution is not a maximum
   entropy state but an unstable saddle point. Known stability
   criteria corresponding to the Maxwellian distribution and the
   water-bag distribution are recovered as particular limits of our
   study. In addition, we find a critical point below which the
   homogeneous Lynden-Bell distribution is always stable. We apply
   these results to the situation considered in Antoniazzi et
   al. For a given energy, we find a critical initial
   magnetization above which the homogeneous Lynden-Bell distribution
   ceases to be a maximum entropy state. For an energy $U=0.69$, this
   transition occurs above an initial magnetization $M_{x}=0.897$. In
   that case, the system should reach an inhomogeneous Lynden-Bell
   distribution (most mixed) or an incompletely mixed state (possibly
   fitted by a Tsallis distribution). Thus, our theoretical study
   proves that the dynamics is different for small and large initial
   magnetizations, in agreement with numerical results of Pluchino et
   al. (2004). This new dynamical phase transition may reconcile the
   two communities by showing that they study different regimes.  
 }

   \maketitle
%

\section{Introduction}
\label{sec_introduction}

The dynamics and thermodynamics of systems with long-range
interactions is a topic of active research \cite{dauxois}. Examples
include self-gravitating systems, two-dimensional vortices, neutral
and non-neutral plasmas, chemotaxis of bacterial populations, just to
mention a few. In addition to these physical systems, a toy model
called the Hamiltonian Mean Field (HMF) model is widely studied [2-21]
because it displays many features of more realistic systems with
long-range interactions, like gravity, while being amenable to a
simpler mathematical treatment \cite{cvb}.

The HMF model is known to display two successive types of relaxation,
like for stellar systems and two dimensional vortices
\cite{houches,pa1}. The first stage of the dynamics is a violent
collisionless relaxation leading to a QSS after a few dynamical
times. This QSS is in general different from the Boltzmann
distribution. The second stage is a slow ``collisional'' relaxation
(due to granularities and finite $N$ effects) leading to the
Boltzmann distribution which is the statistical equilibrium state of
the system. The ``collisional'' relaxation time increases
algebraically with the number of particles so that the QSS has a very
long lifetime which becomes infinite in a proper thermodynamic limit
$N\rightarrow +\infty$.

The nature of the QSS has created an intense debate in the statistical
mechanics community. Two different approaches have been
developed. Some authors \cite{latora,plr,rp} inspired by the work of Tsallis
\cite{tsallis} have proposed to interprete these QSS in terms of a
non-extensive thermodynamics based on the so-called $q$-entropy which
is a generalization of the Boltzmann entropy. Other authors
\cite{yamaguchi,cvb,next05,anto} inspired by the work of Lynden-Bell
\cite{lb,houches} in astrophysics, have proposed to interprete these QSS in
terms of a statistical mechanics of the Vlasov equation called the
theory of violent relaxation \footnote{This theory is relatively well-known in
astrophysics and 2D turbulence \cite{lb,csr,houches} but it took some time to
diffuse in the statistical mechanics community despite several efforts
of the author to publicize it
\cite{thesis,dubrovnik,houches,next05}. In particular, the possibility
to apply the Lynden-Bell theory to the HMF model (but also the limitations
of its application) was mentioned in several papers
\cite{cvb,next05,super}.}. The idea of Lynden-Bell is to determine the
{\it most probable state} of the system resulting from phase mixing
compatible with all the constraints imposed by the Vlasov
dynamics. This assumes that the system mixes well so that a hypothesis
of ergodicity is made. If the initial distribution takes only two
values, Lynden-Bell predicts at meta-equilibrium a coarse-grained
distribution similar to the Fermi-Dirac statistics in quantum
mechanics
\cite{lb,csr,csmnras}. In a recent paper, Antoniazzi et
al. \cite{anto} have performed direct numerical simulations of the HMF
model and found situations in which the Lynden-Bell distribution
provides a good description of the QSS without ad hoc fitting
parameter. These numerical results are a good motivation to
investigate these distributions in more detail, as done in this paper.

In Sec. \ref{sec_vr}, we briefly recall the Lynden-Bell theory of
violent relaxation with notations appropriate to the HMF model. A more
thorough description of this theory can be found in the classical
paper \cite{lb} and in \cite{csr,dubrovnik,super}. In
Sec. \ref{sec_prop}, we determine the equation of state $p(\rho)$
associated with the Lynden-Bell distribution $f(\theta,v)$. We
consider two particular limits: the dilute limit where the Lynden-Bell
distribution becomes similar to the Maxwell-Boltzmann distribution and
the completely degenerate limit where the Lynden-Bell distribution
becomes a step function (water-bag) similar to the Fermi distribution
in quantum mechanics. In Sec. \ref{sec_cc}, we determine the caloric
curve $\beta(E)$ corresponding to the Lynden-Bell statistics in
relation with the HMF model. We restrict ourselves to spatially
homogeneous distributions. In Sec. \ref{sec_stab}, we analyze the
dynamical and thermodynamical stability of these homogeneous solutions
by using two general criteria previously introduced in the literature:
one is based on the distribution function of the system
\cite{yamaguchi} and the other on the velocity of sound
$c_{s}^{2}=p'(\rho)$ in the corresponding barotropic gas
\cite{cvb}. These criteria are equivalent.  We express the critical
energy and the critical temperature as a function of a degeneracy
parameter $\mu$ fixed by the initial condition. The known stability
criteria corresponding to the Maxwell distribution and the water-bag
distribution \cite{yamaguchi,cvb} are recovered as particular cases of
our study since these distributions are two limits of the Lynden-Bell
distribution. For $E<E_{c}(\mu)$ or $T<T_{c}(\mu)$ the homogeneous
Lynden-Bell distribution is {\it not} a maximum entropy
state. Therefore, it is not expected to be achieved as a result of
violent relaxation. In that case, the system may reach an
inhomogeneous Lynden-Bell distribution (if it mixes well), or another
distribution (if it does not mix well). A critical point
$\mu_{*}=0.68786...$ is found below which the homogeneous phase is
always stable whatever the value of energy and temperature. In
Sec. \ref{sec_app}, we apply our results to the situation considered
in Antoniazzi et al. \cite{anto}. For a given energy, we find a
critical initial magnetization above which the homogeneous Lynden-Bell
distribution ceases to be a maximum entropy state and becomes a saddle
point.  For an energy $U=0.69$, this transition occurs above an
initial magnetization $M_{x}=0.897$.  For $M_{x}>0.897$, the system
should reach an inhomogeneous Lynden-Bell distribution (most mixed
state) or an incompletely mixed state (possibly spatially
homogeneous).  Thus, our theoretical study proves that the dynamics is
radically different for small and large initial magnetizations. This
tends to corroborate the claim of Pulchino et al
\cite{plr} who made a similar observation on the basis of numerical
simulations. In the Conclusion, we stress the limitations of the
Lynden-Bell theory and the possibility that the QSS can be described
by other types of distributions when the system does not mix well
(incomplete violent relaxation). Indeed, the Vlasov equation admits an
infinite number of steady states and the system can be trapped in one
of them during the collisionless dynamics \cite{next05}. The Tsallis
distributions (corresponding to stellar polytropes in astrophysics)
are {\it particular} stationary solutions of the Vlasov equation which
can sometimes provide a good fit of the QSS in case of incomplete
relaxation \cite{latora}. However, there is no fundamental reason why
these distributions should always (universally) be selected by the
dynamics and, indeed, many other distributions can emerge in case of
incomplete relaxation, depending on the initial conditions, on the
value of the control parameters, and on the route to equilibrium
\cite{next05}. The Tsallis distributions form just a 
one-parameter family of steady states of the Vlasov equation
\cite{pre} and there is no theoretical justification of using them
unless one invokes their simplicity and popularity. Similarly, stellar
polytropes in astrophysics represent famous stationary solutions of
the Vlasov-Poisson system that can provide simple mathematical models
of galaxies or convenient fits of astrophysical systems in certain
cases, but other distributions can also be considered \cite{bt}. In
fact, real galaxies are {\it not} described by polytropic (or Tsallis)
distributions \cite{aaa,next05}.

\section{Theory of violent relaxation for the HMF model}
\label{sec_vr}

The HMF model is a system of $N$ particles moving on a circle and
interacting via a cosine binary potential, e.g. \cite{ruffo,cvb}. The
dynamics of these particles is governed by the Hamilton equations
\begin{eqnarray}
\label{vr1}
\frac{d\theta_{i}}{dt}=\frac{\partial H}{\partial v_{i}}, \qquad \frac{d v_{i}}{dt}=-\frac{\partial H}{\partial \theta_{i}},\nonumber\\
H=\frac{1}{2}\sum_{i=1}^{N}v_{i}^{2}-\frac{k}{4\pi}\sum_{i\neq j}\cos(\theta_{i}-\theta_{j}).
\end{eqnarray}
This system has an unusual thermodynamic limit defined by
$N\rightarrow +\infty$ with $\epsilon=8\pi E/kM^{2}$ and $\eta=kM/4\pi
T$ fixed (here $M=Nm$ is the total mass and we have taken $m=1$). We
can rescale the parameters of the problem so that the coupling
constant scales like $k\sim 1/N$ while $E\sim N$ and $T\sim 1$ \cite{pa1}. For
$N\rightarrow +\infty$ in this proper thermodynamic limit, the
evolution of the distribution function (DF) is governed by the Vlasov
equation
\cite{yamaguchi,cvb,pa1}:
\begin{eqnarray}
\label{vr2}
\frac{\partial f}{\partial t}+v\frac{\partial f}{\partial\theta}-\frac{\partial\Phi}{\partial\theta}\frac{\partial f}{\partial v}=0,
\end{eqnarray} 
\begin{eqnarray}
\label{vr3}
\Phi(\theta,t)=-\frac{k}{2\pi}\int_{0}^{2\pi}\cos(\theta-\theta')\rho(\theta',t)d\theta'.
\end{eqnarray} 
Starting from an unstable initial DF $f_{0}(\theta,v)$, the Vlasov
equation coupled to the meanfield potential (\ref{vr3}) generates a
complicated mixing process at the end of which the {\it
coarse-grained} DF $\overline{f}(\theta,v,t)$ achieves a
quasi-stationary state $\overline{f}_{QSS}(\theta,v)$. If the system
mixes well \footnote{Lynden-Bell \cite{lb} introduces the notion of
{\it microstates} corresponding to the finely striated structure of
the DF and {\it macrostates} corresponding to the smoothed-out
(coarse-grained) DF. Using the standard postulate of statistical
mechanics, he assumes that all the accessible microstates (with the
right value of the integral constraints) are {\it
equiprobable}. Therefore, if the system ``mixes well'', it will be
found at meta-equilibrium in the macrostate which is represented by
the maximal number of microstates. This {\it most probable} (most
mixed) macrostate is obtained by maximizing the Lynden-Bell mixing
entropy under all the constraints of the Vlasov equation
\cite{super}.}, the QSS is described by the Lynden-Bell distribution
\cite{lb,houches,next05}. If the initial DF takes only two values $f_{0}=\eta_{0}$
and $f=0$, the QSS predicted by Lynden-Bell is obtained by maximizing
the mixing entropy \cite{dubrovnik}:
\begin{eqnarray}
\label{vr4}
S_{LB}=-\int \left \lbrace \frac{\overline{f}}{\eta_{0}}\ln\frac{\overline{f}}{\eta_{0}}+\left (1-\frac{\overline{f}}{\eta_{0}}\right )\ln \left (1-\frac{\overline{f}}{\eta_{0}}\right )\right \rbrace d\theta dv, \nonumber\\
\end{eqnarray} 
at fixed mass 
\begin{eqnarray}
\label{vr5}
M=\int \rho d\theta,
\end{eqnarray} 
and energy
\begin{eqnarray}
\label{vr6}
E=\frac{1}{2}\int f v^{2}d\theta dv+\frac{1}{2}\int \rho\Phi d\theta,
\end{eqnarray} 
where $\rho=\int f dv$ is the spatial density. We thus have to solve the optimization problem
\begin{eqnarray}
\label{vr7}
{\rm Max}\lbrace S[\overline{f}]\quad |\quad E[\overline{f}]=E, M[\overline{f}]=M\rbrace.
\end{eqnarray} 
Writing the first order variations as
\begin{eqnarray}
\label{vr8}
\delta S-\beta\delta E-\alpha\delta M=0
\end{eqnarray} 
where $\beta=1/T$ and $\alpha$ are Lagrange multipliers, one obtains the Lynden-Bell distribution function
\begin{eqnarray}
\label{vr9}
\overline{f}=\frac{\eta_{0}}{1+\lambda e^{\beta (\frac{v^{2}}{2}+\Phi(\theta))}},
\end{eqnarray}
where $\lambda=e^{\alpha}>0$ plays the role of a
fugacity. Morphologically, this distribution function is similar to
the Fermi-Dirac statistics \cite{lb,csr,csmnras} so that we shall find many
analogies with quantum mechanics.

The Lynden-Bell functional (\ref{vr4}) is an entropy because it is
proportional to the logarithm of the disorder, where the disorder is
equal to the number of microstates consistent with a given
macrostate. Indeed, the Lynden-Bell entropy is obtained from a
combinatorial analysis \cite{lb,super}. Therefore, its maximization at
fixed mass and energy determines the most probable macrostate,
i.e. the one that is the most represented at the microscopic
(fine-grained) scale. In this sense, the optimization problem (\ref{vr7}) is a
condition of {\it thermodynamical stability} for the collisionless
relaxation. Alternatively, the Lynden-Bell functional (\ref{vr4}) can also be
intepreted as a particular $H$-function in the sense of Tremaine et
al. \cite{tremaine}. Indeed, it is of the form
\begin{eqnarray}
\label{vr10}
S=-\int C(\overline{f})d\theta dv,
\end{eqnarray} 
where $C$ is convex. In that context, the optimization problem
(\ref{vr7}) is a condition of formal nonlinear dynamical stability
with respect to the Vlasov equation
\cite{tremaine,yamaguchi,cvb,cstsallis,aaantonov}. Therefore, the maximization
of $S$ at fixed $E$ and $M$ guarantees that the statistical
equilibrium macrostate is stable with respect to the perturbation on
the microscopic scale (thermodynamical stability) and that the
coarse-grained DF $\overline{f}$ is stable for the Vlasov equation
with respect to macroscopic perturbations (nonlinear dynamical
stability).

We emphasize that it is only when the initial DF takes two values
$\eta_{0}$ and $0$ that the Lynden-Bell entropy can be expressed in
terms of the coarse-grained DF $\overline{f}$ as in Eq. (\ref{vr4}). In
general, the Lynden-Bell entropy is a functional of the probability
distribution of phase levels $\rho(\theta,v,\eta)$ of the form:
\begin{eqnarray}
\label{vr11}
S_{LB}[\rho]=-\int \rho\ln\rho \, d\theta dv d\eta,
\end{eqnarray}    
and the coarse-grained DF is given by $\overline{f}=\int \rho\eta
d\eta$ \cite{lb,csr,super,next05}. The general Lynden-Bell distribution is
expressed as a superposition of Fermi-Dirac distributions of the form
\begin{eqnarray}
\label{vr12}
\overline{f}=\frac{\int \chi(\eta)\eta e^{-\eta(\beta\epsilon+\alpha)}d\eta}{\int \chi(\eta) e^{-\eta(\beta\epsilon+\alpha)}d\eta},
\end{eqnarray} 
where we have noted $\epsilon=v^{2}/2+\Phi(\theta)$ the energy per
particle. This is similar to a sort of superstatistics
\cite{super} where the function $\chi(\eta)$ is determined indirectly by the initial condition. Therefore, the expression (\ref{vr4}) of the
collisionless entropy is not universal; it is valid only in the
two-levels approximation. We note also that the expression of the
collisionless entropy given by \cite{rf} is not correct. These authors
do not introduce the notion of coarse-graining and phase mixing, nor
the local distribution of phase levels $\rho(\theta,v,\eta)$, which is
capital in the theory of violent relaxation to describe the QSS \cite{lb}. The
correct form of entropy for the violent relaxation process (based on
the Vlasov equation) is the Lynden-Bell entropy (\ref{vr11}) as
claimed in
\cite{next05}.

It should be stressed that the theory of violent relaxation is valid
for many systems with long-range interactions described by the Vlasov
equation, not only for the HMF model \cite{pa1}. Historically, this theory was
first introduced in astrophysics for collisionless stellar systems
\cite{lb}. The calculation of self-gravitating 
Fermi-Dirac spheres corresponding to the
Lynden-Bell distribution was performed in \cite{csmnras} and the
caloric curve $\beta(E)$ was obtained as a function of a degeneracy
parameter $\mu$. An equivalent theory of violent relaxation was
developed in two-dimensional turbulence described by the 2D Euler
equation to account for the structure and robustness of large-scale
vortices such as Jupiter's great red spot \cite{miller,rs}. The
analogy between 2D vortices and stellar systems was discussed in
\cite{thesis,csr,houches}. Many numerical simulations have been performed in
the two domains to test the successes and the failures of the
Lynden-Bell prediction (see \cite{next05} for some references). The
limitations of the Lynden-Bell theory will be discussed in the
Conclusion. In the following, we shall assume that the QSS is
described by the Lynden-Bell distribution (statistical equilibrium
state of the Vlasov equation) and we use a presentation similar to
that developed in the gravitational context
\cite{csmnras}.

\section{Properties of the Lynden-Bell distribution}
\label{sec_prop}

In this section, we discuss some properties of the Lynden-Bell
distribution (\ref{vr9}) and consider two limit forms of this
distribution: the Maxwell-Boltzmann distribution obtained in the
non-degenerate limit and the Fermi distribution (water-bag) obtained
in the completely degenerate limit.

\subsection{The equation of state}
\label{sec_es}

To any distribution function $\overline{f}=\overline{f}(\epsilon)$
depending only on the energy $\epsilon=v^{2}/2+\Phi(\theta)$, one can
associate a corresponding barotropic equation of state $p(\rho)$
\cite{cvb}. The density and the pressure are defined by
\begin{eqnarray}
\label{es1}
\rho=\int_{-\infty}^{+\infty} \overline{f} dv,
\end{eqnarray}
\begin{eqnarray}
\label{es2}
p=\int_{-\infty}^{+\infty} \overline{f} v^{2} dv.
\end{eqnarray}
Let us determine the equation of state associated with the Lynden-Bell
distribution (\ref{vr9}). Substituting for $\overline{f}$ from
Eq. (\ref{vr9}) to Eqs. (\ref{es1})-(\ref{es2}), introducing the
notation $\Lambda(\theta)=\lambda e^{\beta\Phi(\theta)}$, and
performing the change of variables $x=\beta v^{2}/2$, we obtain
\begin{eqnarray}
\label{es3}
\rho=\left (\frac{2}{\beta}\right )^{1/2}\eta_{0}I_{-1/2}(\Lambda),
\end{eqnarray}
\begin{eqnarray}
\label{es4}
p=\left (\frac{2}{\beta}\right )^{3/2}\eta_{0}I_{1/2}(\Lambda),
\end{eqnarray}
where we have defined the Fermi integrals 
\begin{eqnarray}
\label{es5}
I_{n}(t)=\int_{0}^{+\infty}\frac{x^{n}}{1+t e^{x}}dx.
\end{eqnarray}
By eliminating $\Lambda$ between Eq. (\ref{es3}) and Eq. (\ref{es4}),
we see that the equation of state is barotropic, i.e. the pressure is
a function $p(\rho)$ of the density. It is equivalent to the equation
of state of an ideal Fermi gas in one dimension.

\subsection{The dilute limit (Maxwell-Boltzmann distribution)}
\label{sec_d}

In the limit $\Lambda\rightarrow +\infty$, the Lynden-Bell distribution reduces to the Maxwell-Boltzmann distribution  
\begin{eqnarray}
\label{d1}
\overline{f}\simeq \frac{\eta_{0}}{\lambda} e^{-\beta (\frac{v^{2}}{2}+\Phi(\theta))}.
\end{eqnarray}
Since $\overline{f}\ll \eta_{0}$, this corresponds to a dilute limit
(or to a non degenerate limit if we use the terminology of quantum
mechanics). In this limit, the Lynden-Bell entropy (\ref{vr4}) takes a form
similar to the Boltzmann entropy
\begin{eqnarray}
\label{d2}
S_{LB}\simeq -\int \frac{\overline{f}}{\eta_{0}}\ln\frac{\overline{f}}{\eta_{0}} d\theta dv.
\end{eqnarray}
The corresponding equation of state
is that of an isothermal gas
\begin{eqnarray}
\label{d3}
p=\rho T.
\end{eqnarray}
This result can be obtained directly from Eqs. (\ref{es3})-(\ref{es4})
by using the asymptotic expression of the Fermi-Dirac integrals for
$t\rightarrow +\infty$:
\begin{eqnarray}
\label{d4}
I_{n}(t)\sim\frac{\Gamma(n+1)}{t}, \qquad (n>-1).
\end{eqnarray}

\subsection{The degenerate limit (water-bag distribution)}
\label{sec_deg}

In the limit $\Lambda\rightarrow 0$, the Lynden-Bell distribution (\ref{vr9}) reduces to the  Heaviside function
\begin{eqnarray}
\overline{f}=\Biggl\lbrace \begin{array}{cc}
\eta_{0}   & \qquad (v< v_{F}), \\
0  & \qquad (v\ge v_{F}),
\end{array}
\label{deg1}
\end{eqnarray}
where 
\begin{eqnarray}
\label{deg2}
v_{F}(\theta)=\sqrt{-(2/\beta)\ln\Lambda(\theta)},
\end{eqnarray}
is a maximum velocity.  The distribution (\ref{deg1}) is often called the
water-bag distribution. It is also similar to the Fermi distribution
in quantum mechanics and $v_{F}$ is similar to the Fermi
velocity. Thus, the limit $\Lambda\rightarrow 0$ corresponds to a
completely degenerate limit in the quantum mechanics terminology.

Using Eqs. (\ref{es1})-(\ref{es2}), the density is given by
$\rho=2\eta_{0}v_{F}$ and the pressure by
$p=(2/3)\eta_{0}v_{F}^{3}$. Eliminating $v_{F}$ between these two
expressions, we find that the equation of state is
\begin{eqnarray}
\label{deg3}
p=\frac{1}{12\eta_{0}^{2}}\rho^{3}.
\end{eqnarray}
This is similar to the equation of state of a polytrope
$p=K\rho^{1+1/n}$ with an index $n=1/2$ and a polytropic constant
$K=1/(12\eta_{0}^{2})$. Polytropic distributions (related to Tsallis
distributions) have been studied in \cite{cvb} in relation with the
HMF model. The equation of state (\ref{deg3}) can also be obtained directly
from Eqs. (\ref{es3})-(\ref{es4})   by using the asymptotic expression of the
Fermi-Dirac integrals for $t\rightarrow 0$:
\begin{eqnarray}
\label{deg4}
I_{n}(t)\sim\frac{(-\ln t)^{n+1}}{n+1}, \qquad (n>-1).
\end{eqnarray}

\section{The caloric curve}
\label{sec_cc}

From now on, we restrict ourselves to spatially homogeneous systems ($\Phi=0$) so that the Lynden-Bell distribution becomes
\begin{eqnarray}
\label{cc1}
\overline{f}=\frac{\eta_{0}}{1+\lambda e^{\beta \frac{v^{2}}{2}}}.
\end{eqnarray}
In this section, we shall determine the relation between the
temperature $T$ and the energy $E$. This defines the caloric
curve $T(E)$. Note that the temperature $T$ is a Lagrange multiplier
associated with the conservation of energy in the variational problem
(\ref{vr8}). It also has the interpretation of a kinetic temperature in the
Fermi-Dirac distribution (\ref{vr9}).

\begin{figure}
\centering
\includegraphics[width=8cm]{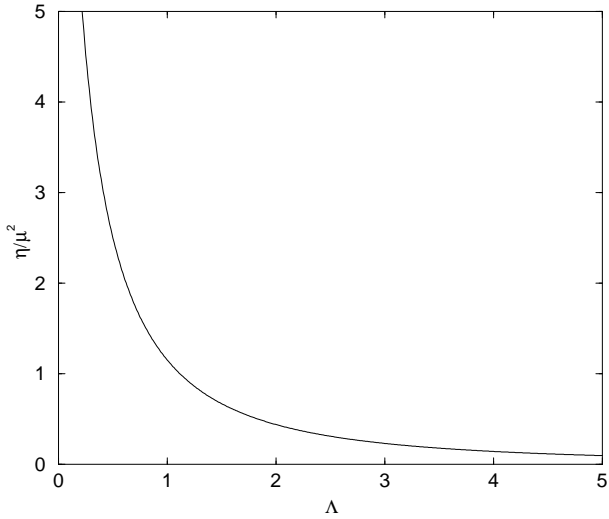}
\caption{Inverse temperature $\eta$ as a function of the fugacity $\Lambda$.}
\label{etalambdaBON}
\end{figure}

It is useful to introduce dimensionless quantities as in
\cite{cvb}. We define the dimensionless inverse temperature by
\begin{eqnarray}
\label{cc2}
\eta=\frac{kM}{4\pi T}.
\end{eqnarray}
We also introduce the degeneracy parameter
\begin{eqnarray}
\label{cc3}
\mu=\eta_{0}\left (\frac{2\pi k}{M}\right )^{1/2}.
\end{eqnarray}
These notations are similar to those used in the astrophysical context
\cite{csmnras}. Using Eq. (\ref{es3}) and $\rho=M/(2\pi)$ for a homogeneous
system, we find that the parameter $\Lambda$ (fugacity) is related to the
temperature and to the degeneracy parameter by
\begin{eqnarray}
\label{cc4}
\eta=\mu^{2}I_{-1/2}(\Lambda)^{2}.
\end{eqnarray}
The curve $\eta(\Lambda)$ is decreasing. It behaves as $\eta\sim -4\mu^{2}\ln\Lambda$ for $\Lambda\rightarrow 0$ and as $\eta\sim \pi\mu^{2}/\Lambda^{2}$ for $\Lambda\rightarrow +\infty$ (see Fig. \ref{etalambdaBON}).

\begin{figure}
\centering
\includegraphics[width=8cm]{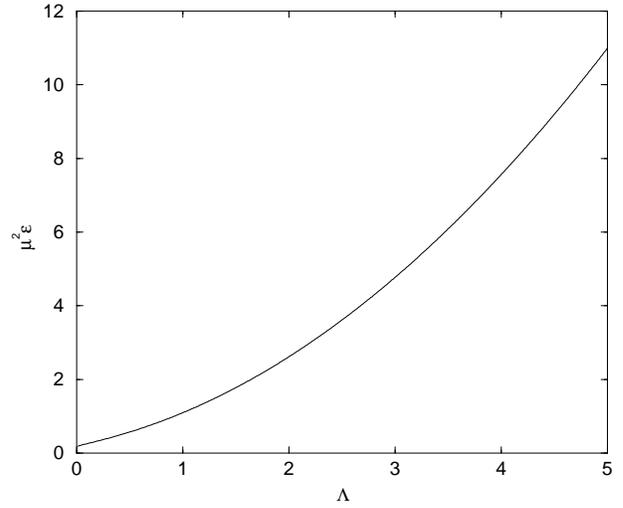}
\caption{Energy $\epsilon$ as a function of the fugacity $\Lambda$. }
\label{epsilonlambdaBON}
\end{figure}

For a homogeneous system, the energy is simply the kinetic energy
\begin{eqnarray}
\label{cc5}
E=\frac{1}{2}\int \overline{f} v^{2}dv d\theta.
\end{eqnarray}
In terms of the pressure (\ref{es2}), this can be written
\begin{eqnarray}
\label{cc6}
E=\pi p.
\end{eqnarray}
Introducing the dimensionless energy \cite{cvb}:
\begin{eqnarray}
\label{cc7}
\epsilon=\frac{8\pi E}{kM^{2}},
\end{eqnarray}
and using Eq. (\ref{es4}), we get
\begin{eqnarray}
\label{cc8}
\epsilon=\frac{2\mu}{\eta^{3/2}}I_{1/2}(\Lambda).
\end{eqnarray}
Using Eq. (\ref{cc4}), the foregoing equation can be rewritten
\begin{eqnarray}
\label{cc9}
\epsilon=\frac{2}{\mu^{2}}\frac{I_{1/2}(\Lambda)}{I_{-1/2}(\Lambda)^{3}}.
\end{eqnarray}
The function $\epsilon(\Lambda)$ increases (see Fig. \ref{epsilonlambdaBON}). It starts from
$\epsilon(0)=1/(6\mu^{2})$ and increases like $\epsilon(\Lambda)\sim
\Lambda^{2}/(\pi\mu^{2})$ for $\Lambda\rightarrow +\infty$. Therefore,
\begin{eqnarray}
\label{cc10}
\epsilon\ge \epsilon_{min}=\frac{1}{6\mu^{2}}.
\end{eqnarray}
This minimum energy is similar to the ground state of a one
dimensional Fermi gas. In terms of dimensional variables it is given
by
\begin{eqnarray}
\label{cc11}
E_{min}=\frac{M^{3}}{96\pi^{2}\eta_{0}^{2}}.
\end{eqnarray}

The caloric curve $\beta(E)$, or equivalently $\eta(\epsilon)$, is
obtained by eliminating $\Lambda$ between Eqs. (\ref{cc4}) and
(\ref{cc9}). We note that the relation between $\eta/\mu^{2}$ and
$\epsilon\mu^{2}$ is independent on the degeneracy parameter $\mu$. For $\Lambda\rightarrow +\infty$, we
recover the relation
\begin{eqnarray}
\label{cc12}
\eta=\frac{1}{\epsilon}, \qquad (\epsilon\rightarrow +\infty),
\end{eqnarray} 
valid for a classical isothermal gas described by the
Maxwell-Boltzmann distribution (\ref{d1}) \cite{cvb}. In terms of
dimensional variables, the relation (\ref{cc12}) can be written
\begin{eqnarray}
\label{cc13}
E=\frac{1}{2}MT, \qquad (E\rightarrow +\infty).
\end{eqnarray} 
To investigate the behaviour of the caloric curve close to the ground
state, we use the Sommerfeld expansion of the Fermi integrals for
$t\rightarrow 0$:
\begin{eqnarray}
\label{cc14}
I_{1/2}(t)=\frac{2}{3}(-\ln t)^{3/2}\left (1+\frac{\pi^{2}}{8}(-\ln t)^{-2}+...\right ),
\end{eqnarray} 
\begin{eqnarray}
\label{cc15}
I_{-1/2}(t)=2(-\ln t)^{1/2}\left (1-\frac{\pi^{2}}{24}(-\ln t)^{-2}+...\right ).
\end{eqnarray} 
Combining these results with Eqs. (\ref{cc4})-(\ref{cc9}), we obtain
\begin{eqnarray}
\label{cc16}
\eta=\left (\frac{2}{3}\right )^{1/2}\pi \mu (\epsilon-\epsilon_{min})^{-1/2}, \qquad (\epsilon\rightarrow \epsilon_{min}).
\end{eqnarray} 
In terms of dimensional variables, this relation can be written
\begin{eqnarray}
\label{cc17}
E=E_{min}\left \lbrack 1+\frac{\pi^{2}}{36}\left (\frac{MT}{E_{min}}\right )^{2}+...\right \rbrack.
\end{eqnarray} 
We note that the energy does not vanish for $T=0$. This is similar to
the effect of a quantum pressure in quantum mechanics, i.e. the distribution function (\ref{cc1}) is {\it not} a Dirac peak $M\delta(v)$ for $T=0$.

\begin{figure}
\centering
\includegraphics[width=8cm]{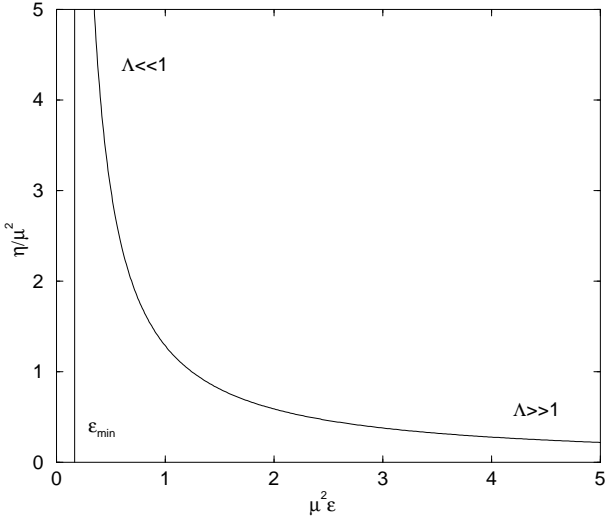}
\caption{Caloric curve corresponding to the spatially homogeneous Lynden-Bell distribution.}
\label{etaepsilon}
\end{figure}

The caloric curve $\eta(\epsilon)$ is represented in
Fig. \ref{etaepsilon}. It is parameterized by $\Lambda$. We
note that the dilute limit $\Lambda\rightarrow +\infty$ corresponds to
$\epsilon\rightarrow +\infty$ and the degenerate limit
$\Lambda\rightarrow 0$ corresponds to $\epsilon\rightarrow
\epsilon_{min}$.

\section{Stability of the homogeneous phase}
\label{sec_stab}

We have seen that the optimization problem (\ref{vr7}) provides a
condition of thermodynamical stability (in Lynden-Bell's sense) and a
condition of nonlinear dynamical stability with respect to the Vlasov
equation. We thus have to select the {\it maximum} of $S$ at fixed
$E$, $M$. Indeed, a saddle point of $S$ is unstable and cannot be
obtained as a result of a violent relaxation. It can be shown that,
for the HMF model, the optimization problem (\ref{vr7}) is equivalent
to the optimization problem
\begin{eqnarray}
\label{stab1}
{\rm Min}\lbrace F[\overline{f}]=E[\overline{f}]-T S[\overline{f}]\quad |\quad  M[\overline{f}]=M\rbrace,
\end{eqnarray} 
where $F$ can be interpreted as a free energy. The criterion
(\ref{vr7}) can be viewed as a criterion of microcanonical stability
and the criterion (\ref{stab1}) as a criterion of canonical
stability. For the HMF model, the statistical ensembles
(microcanonical and canonical) are equivalent so that all the stable
solutions can be constructed from the simpler optimization problem
(\ref{stab1}); no stable solution is forgotten if we solve (\ref{stab1})
instead of (\ref{vr7}). Now, the optimization problem (\ref{stab1})
has been studied in \cite{yamaguchi,cvb} for general functionals of
the form (\ref{vr10}) and a simple stability criterion has been obtained
in the case where the steady state is spatially homogeneous. The
stability criterion can be expressed either in terms of the
distribution function \cite{yamaguchi} or in terms of the velocity of
sound in the corresponding barotropic gas \cite{cvb}. In this section,
we apply these criteria to the Lynden-Bell distribution (\ref{cc1}).

\subsection{Criterion based on the velocity of sound}
\label{sec_vs}

It is shown in \cite{cvb} that the stability criterion (\ref{stab1})
for a spatially homogeneous solution of the Vlasov equation can be
put in the form of a condition on the velocity of sound
$c_{s}^{2}=p'(\rho)$ in the corresponding barotropic gas. A spatially
homogeneous distribution is stable with respect to the Vlasov equation
(in the above sense) if, and only, if
\begin{eqnarray}
\label{vs1}
c_{s}^{2}\ge \frac{kM}{4\pi}.
\end{eqnarray}
This stability criterion exploits the subtle correspondance between a kinetic
system described by a DF $f=f(\epsilon)$ and a barotropic gas with an
equation of state $p(\rho)$. This correspondance is related to the
Antonov first law in astrophysics (see \cite{cvb,aaantonov} for details). 

\begin{figure}
\centering
\includegraphics[width=8cm]{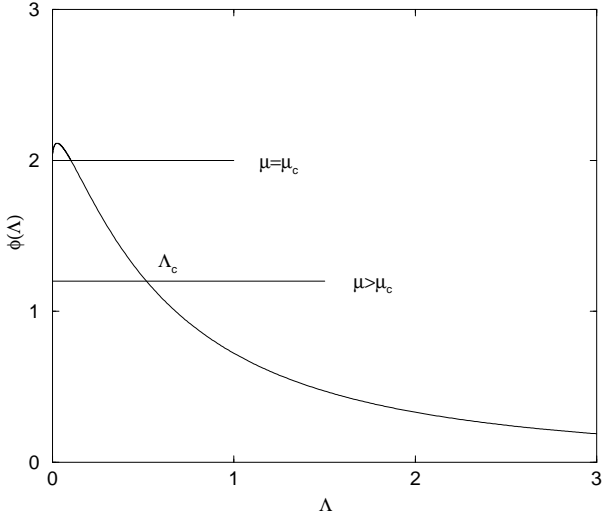}
\caption{Graphical construction determining the critical value of the fugacity $\Lambda_{c}$ below which the homogeneous Lynden-Bell distribution is not a maximum entropy state anymore. For $\mu>\mu_{c}$ there is only one intersection.  }
\label{phiBON}
\end{figure}

\begin{figure}
\centering
\includegraphics[width=8cm]{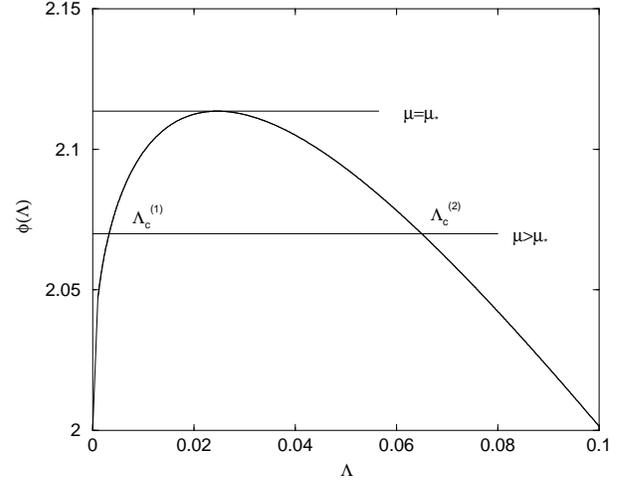}
\caption{Graphical construction determining the critical values of the fugacity $\Lambda_{c}$ at which the homogeneous Lynden-Bell distribution ceases to be a maximum entropy state. For $\mu_{*}<\mu<\mu_{c}$ there are two intersections. The homogeneous distribution is stable (entropy maximum) for $\Lambda<\Lambda_{c}^{(1)}$ and $\Lambda>\Lambda_{c}^{(2)}$. }
\label{phiZOOM}
\end{figure}

Let us now apply this criterion to the Lynden-Bell distribution. From
Eqs. (\ref{es3})-(\ref{es4}), we get
\begin{eqnarray}
\label{vs2}
p'(\rho)=\frac{2}{\beta}\frac{I_{1/2}'(\Lambda)}{I_{-1/2}'(\Lambda)}.
\end{eqnarray}
Now, using the identity
\begin{eqnarray}
\label{vs3}
I_{n}'(t)=-\frac{n}{t}I_{n-1}(t), \qquad (n>0)
\end{eqnarray}
we obtain
\begin{eqnarray}
\label{vs4}
c_{s}^{2}=\frac{1}{\Lambda\beta }\frac{I_{-1/2}(\Lambda)}{|I_{-1/2}'(\Lambda)|}.
\end{eqnarray}
In terms of the dimensionless parameters, the stability criterion (\ref{vs1}) can be written
\begin{eqnarray}
\label{vs5}
\eta\frac{\Lambda |I_{-1/2}'(\Lambda)|}{I_{-1/2}(\Lambda)}\le 1,
\end{eqnarray}
where $\Lambda$ is given by
\begin{eqnarray}
\label{vs6}
I_{-1/2}(\Lambda)=\frac{\sqrt{\eta}}{\mu},
\end{eqnarray}
according to Eq. (\ref{cc4}). Combining Eqs. (\ref{vs5})-(\ref{vs6}), we can rewrite the stability criterion in the form
\begin{eqnarray}
\label{vs7}
\phi(\Lambda)\equiv I_{-1/2}(\Lambda)\Lambda |I_{-1/2}'(\Lambda)|\le \frac{1}{\mu^{2}}.
\end{eqnarray}
The function $\phi(\Lambda)$ starts from $\phi(0)=2$. It first
increases like $\phi(\Lambda)=2+(\pi^{2}/6)(-\ln \Lambda)^{-2}$ for
$\Lambda\rightarrow 0$, reaches a maximum at
$(\Lambda_{*}=0.024,\phi_{*}=2.1135)$ and then decreases like
$\phi(\Lambda)\sim
\pi/\Lambda^{2}$ for $\Lambda\rightarrow +\infty$ (see Figs. \ref{phiBON}-\ref{phiZOOM}). Therefore, there exists a critical point in the problem. If
\begin{eqnarray}
\label{vs8}
\mu\le \mu_{*}\equiv \frac{1}{\sqrt{\phi_{*}}}=0.68786,
\end{eqnarray}
the homogeneous system is stable for any temperature and any
energy. In terms of dimensional quantities, this corresponds to
\begin{eqnarray}
\label{vs9}
 \eta_{0}\le \mu_{*}\left (\frac{M}{2\pi k}\right )^{1/2}.
\end{eqnarray}
Alternatively, for $\mu>\mu_{*}$ the homogeneous Lynden-Bell
distribution is not always a maximum entropy state.

For $\mu>\mu_{c}$, where
\begin{eqnarray}
\label{vs10}
\mu_{c}=\frac{1}{\sqrt{2}}=0.70710,
\end{eqnarray}
the equation $\phi(\Lambda)=1/\mu^{2}$ has only one solution denoted
$\Lambda_{c}$ (see Fig. \ref{phiBON}). The condition of stability of
the homogenous phase corresponds to $\Lambda>\Lambda_{c}$. In terms of
the temperature (\ref{cc4}) or the energy (\ref{cc9}), the condition
of stability of the homogeneous phase can be written
\begin{eqnarray}
\label{vs11}
\eta\le \eta_{c}(\mu), \qquad \epsilon\ge \epsilon_{c}(\mu),
\end{eqnarray}
where the critical temperature and the critical energy are defined by the parametric equations
\begin{eqnarray}
\label{vs12}
\Lambda_{c} I_{-1/2}(\Lambda_{c}) |I_{-1/2}'(\Lambda_{c})|=\frac{1}{\mu^{2}},
\end{eqnarray}
\begin{eqnarray}
\label{vs13}
\eta_{c}=\frac{I_{-1/2}(\Lambda_{c})}{\Lambda_{c} |I_{-1/2}'(\Lambda_{c})|},
\end{eqnarray}
\begin{eqnarray}
\label{vs14}
\epsilon_{c}=\frac{2 \Lambda_{c}  |I_{-1/2}'(\Lambda_{c})|}{I_{-1/2}(\Lambda_{c})^{2}}I_{1/2}(\Lambda_{c}),
\end{eqnarray}
where we recall that
\begin{eqnarray}
\label{vs15}
I_{-1/2}(t)=\int_{0}^{+\infty}\frac{1}{\sqrt{x}(1+t e^{x})}dx,
\end{eqnarray}
\begin{eqnarray}
\label{vs16}
I_{-1/2}'(t)=-\int_{0}^{+\infty}\frac{e^{x}}{\sqrt{x}(1+t e^{x})^2}dx,
\end{eqnarray}
\begin{eqnarray}
\label{vs17}
I_{1/2}(t)=\int_{0}^{+\infty}\frac{\sqrt{x}}{1+t e^{x}}dx.
\end{eqnarray}

For $\mu_{*}<\mu<\mu_{c}$, the equation $\phi(\Lambda)=1/\mu^{2}$ has
two solutions denoted $\Lambda_{c}^{(1)}$ and $\Lambda_{c}^{(2)}$ (see
Fig. \ref{phiZOOM}). The homogeneous phase is stable for
$\Lambda<\Lambda_{c}^{(1)}$ and for $\Lambda>\Lambda_{c}^{(2)}$. In
terms of the temperature, the condition of stability of the
homogeneous distribution can be written
\begin{eqnarray}
\label{vs18}
\eta\le \eta_{c}^{(2)}(\mu)\qquad {\rm or} \qquad \eta\ge \eta_{c}^{(1)}(\mu),
\end{eqnarray}
where $\eta_{c}^{(1)}(\mu)$ and $\eta_{c}^{(2)}(\mu)$ are given by
Eq. (\ref{vs13}) with $\Lambda_{c}^{(1)}$ and $\Lambda_{c}^{(2)}$
respectively. In terms of the energy, the condition of stability of
the homogeneous distribution can be written
\begin{eqnarray}
\label{vs19}
\epsilon\ge \epsilon_{c}^{(2)}(\mu)\qquad {\rm or} \qquad \epsilon_{min}(\mu)\le \epsilon\le \epsilon_{c}^{(1)}(\mu),
\end{eqnarray}
where $\epsilon_{c}^{(1)}(\mu)$ and $\epsilon_{c}^{(2)}(\mu)$ are
given by Eq. (\ref{vs13}) with $\Lambda_{c}^{(1)}$ and
$\Lambda_{c}^{(2)}$ respectively.

\begin{figure}
\centering
\includegraphics[width=8cm]{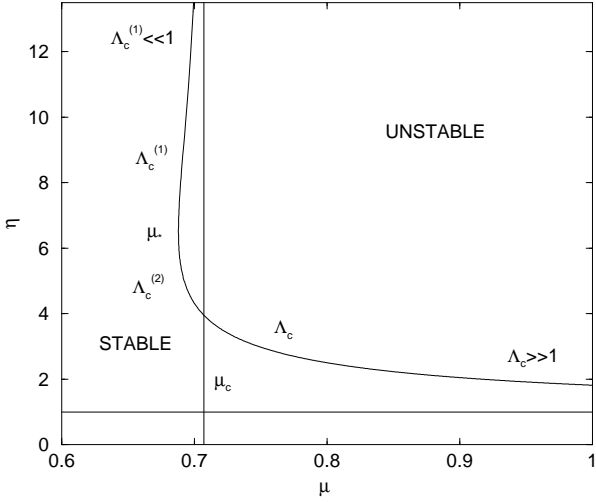}
\caption{Stability diagram of the homogeneous phase in the $(\mu,\eta)$ plane.}
\label{etamu}
\end{figure}

\begin{figure}
\centering
\includegraphics[width=8cm]{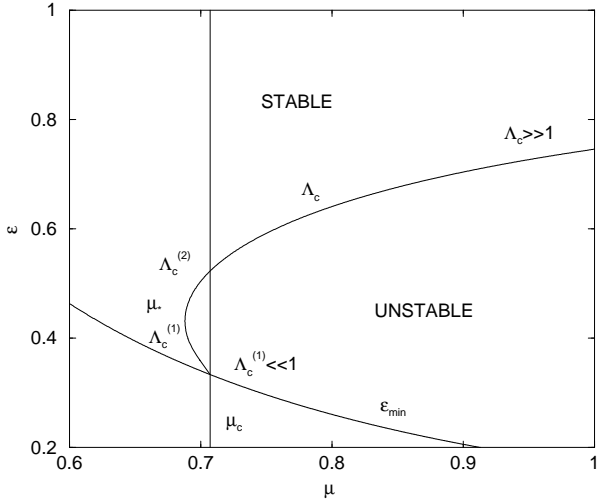}
\caption{Stability diagram of the homogeneous phase in the $(\mu,\epsilon)$ plane.}
\label{muepsilon}
\end{figure}

\begin{figure}
\centering
\includegraphics[width=8cm]{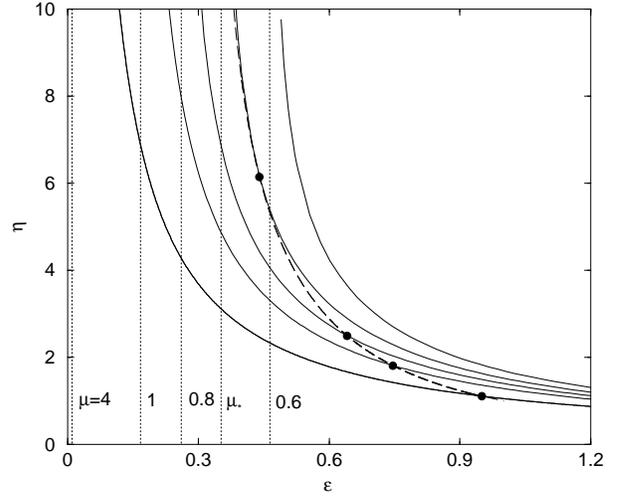}
\caption{Caloric curve (\ref{cc4})-(\ref{cc9}) corresponding to the homogeneous Lynden-Bell distribution for different values of the degeneracy parameter $\mu$. We have indicated the point $(\epsilon_{c},\eta_{c})$ at which the series of equilibria becomes unstable. These points are related to each other by the dashed line $(\epsilon_{c}(\mu)-\eta_{c}(\mu))$ parameterized by $\mu$. It is obtained from Eqs. (\ref{vs13})-(\ref{vs14}). For $\mu>\mu_{c}$ there is only one intersection between the caloric curve and the dashed line. For $\mu_{*}<\mu<\mu_{c}$ there are two intersections (see Fig. \ref{stabZOOM}). For $\mu<\mu_{*}$ there is no intersection and the homogeneous phase is always stable. }
\label{stab}
\end{figure}

\begin{figure}
\centering
\includegraphics[width=8cm]{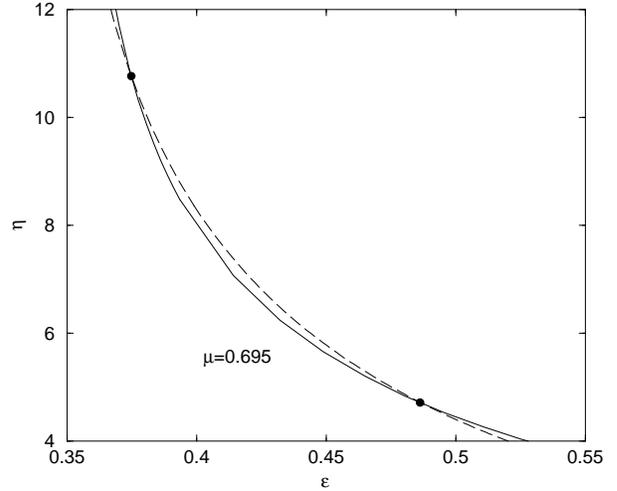}
\caption{Enlargement of the previous diagram to show the particularity of the interval $\mu_{*}<\mu<\mu_{c}$. When $\mu$ lies in this interval  (specifically we have taken $\mu=0.695$) there exists two zones of stability in the series of equilibria separated by a zone of instability.}
\label{stabZOOM}
\end{figure}

The stability diagram of the homogeneous Lynden-Bell distribution
(\ref{cc1}) is plotted in Figs.  \ref{etamu}-\ref{muepsilon} in the
$(\mu,\eta)$ plane and in the $(\mu,\epsilon)$ plane respectively.
The representative curve $\eta_{c}(\mu)$ or $\epsilon_{c}(\mu)$ marks
the separation between the stable (maximum entropy states) and the
unstable (saddles point of entropy) regions.  We have also plotted the
minimum accessible energy $\epsilon_{min}(\mu)$.  In
Figs. \ref{stab}-\ref{stabZOOM}, we have represented the caloric curve
$\eta(\epsilon)$ for different values of the degeneracy parameter and
we have indicated the point at which the series of equilibria becomes
unstable (for sufficiently small values of $\epsilon$ or sufficiently
large values of $\eta$). Here, the term ``unstable'' means that the
homogeneous Lynden-Bell distribution is not a maximum entropy state,
i.e. (i) it is not the most mixed state (ii) it is dynamically
unstable with respect to the Vlasov equation. As a result, it should
not be reached as a result of violent relaxation. One possibility is
that the system converges to the spatially {\it inhomogeneous}
Lynden-Bell distribution (\ref{vr9}) with $\Phi\neq 0$ which is the
maximum entropy state (most mixed) in that case (see
Sec. \ref{sec_app}). Another possibility, always to consider, is that
the system does not converge towards the maximum entropy state,
i.e. the relaxation is {\it incomplete} (see Sec. \ref{sec_app} and
the Conclusion).

\subsection{Limit cases}
\label{sec_lc}

In the non-degenerate limit $\Lambda_{c}\rightarrow +\infty$, using
the asymptotic expansion (\ref{d4}), we find that
\begin{eqnarray}
\label{lc1}
\eta_{c}\rightarrow 1, \qquad \epsilon_{c}\rightarrow 1, \qquad \mu\sim \Lambda_{c}/\sqrt{\pi}.
\end{eqnarray}
These results are valid for $\mu\rightarrow +\infty$. In terms of
dimensional variables, the condition of stability can be rewritten
\begin{eqnarray}
\label{lc2}
T\ge \frac{kM}{4\pi}\equiv T_{c}, \qquad  E\ge \frac{kM^{2}}{8\pi}\equiv E_{c}.
\end{eqnarray}
This returns the well-known nonlinear dynamical stability criterion
(with respect to the Vlasov equation) of a homogeneous system with
Maxwellian distribution function (see, e.g.,
\cite{yamaguchi,cvb}). Indeed, for the equation of state (\ref{d3}),
the velocity of sound is $c_{s}^{2}=T$ and the stability criterion
(\ref{vs1}) directly leads to Eq. (\ref{lc2}). This also coincides
with the ordinary thermodynamical stability criterion applying to the
{\it collisional} regime, for $t\rightarrow +\infty$, where the
statistical equilibrium state is the Boltzmann distribution for $f$
(without the bar!).

In the completely degenerate limit $\Lambda_{c}\rightarrow 0$, using the asymptotic expansion (\ref{deg4}), we find that 
\begin{eqnarray}
\label{lc3}
\eta_{c}\rightarrow +\infty, \qquad \epsilon_{c}\rightarrow \frac{1}{3}, \qquad \mu\rightarrow \frac{1}{\sqrt{2}}.
\end{eqnarray}
This can be interpreted in terms of the water-bag model described by
the DF (\ref{deg1}).  For the equation of state (\ref{deg3}), the
velocity of sound is
$c_{s}^{2}=\rho^{2}/(4\eta_{0}^{2})=M^{2}/(16\pi^{2}\eta_{0}^{2})$ and
the stability criterion (\ref{vs1}) gives
\begin{eqnarray}
\label{lc4}
\eta_{0}\le \left (\frac{M}{4\pi k}\right )^{1/2}, \qquad {\rm i.e.} \qquad \mu\le \frac{1}{\sqrt{2}}.
\end{eqnarray}
Note, parenthetically, that this criterion can also be written as a
condition on the Fermi velocity \cite{cvb}: $v_{F}^{2}\ge kM/(4\pi)$
since $c_{s}=v_{F}$ according to the relations of
Sec. \ref{sec_deg}. For any value of $\mu$, the Fermi distribution is
valid for the energy $\epsilon=\epsilon_{min}=1/(6\mu^{2})$
corresponding to $\Lambda\rightarrow 0$ (ground state). According to
the criterion (\ref{lc4}), it is stable only for $\mu\le
\mu_{c}=1/\sqrt{2}$, i.e
\begin{eqnarray}
\label{lc5}
\epsilon_{min}\ge \frac{1}{3}. 
\end{eqnarray}
This returns the well-known stability criterion for the water-bag
model (see, e.g., \cite{yamaguchi,cvb}).

\subsection{Criterion based on the distribution function}
\label{sec_cd}

As shown in \cite{yamaguchi}, the stability criterion associated with
the optimization problem (\ref{stab1}) can be written in terms of the
distribution function as
\begin{eqnarray}
\label{cd1}
1+\frac{k}{2}\int_{-\infty}^{+\infty}\frac{f'(v)}{v}dv\le 0.
\end{eqnarray}
The equivalence with the criterion (\ref{vs1}) is proved in
\cite{cvb}. The criterion (\ref{cd1}) can also be obtained by investigating the {\it linear} dynamical stability of a homogeneous solution of the Vlasov equation \cite{inagaki,choi,cvb}. Substituting Eq. (\ref{cc1}) in Eq. (\ref{cd1}), we find that a
homogeneous system described by the Lynden-Bell distribution is stable
if and only if
\begin{eqnarray}
\label{cd2}
1-k\eta_{0}\Lambda \left (\frac{\beta}{2}\right )^{1/2}\int_{0}^{+\infty}\frac{e^{x}}{\sqrt{x}(1+\Lambda e^{x})^{2}}dx\ge 0.
\end{eqnarray}
This condition is equivalent to Eq. (\ref{vs5}), as it should, so that
the previous stability analysis could have been performed without
modification by starting directly from Eq. (\ref{cd1}).

We may note at this place that the notations introduced in this paper
and in \cite{cvb} differ from those usually introduced in the HMF
literature, e.g. \cite{yamaguchi}. This is because we tried to draw a
close parallel with the notations introduced in astrophysics,
e.g. \cite{aa1}. However, it is not difficult to find the relation
between the two sets of parameters. In particular, we have
\begin{eqnarray}
\label{cd3}
\epsilon=4\left ( U-\frac{1}{2}\right ), \qquad \eta=\frac{\beta}{2},
\end{eqnarray}
where $U$ and $\beta$ are the energy and the inverse temperature used
in \cite{yamaguchi} (note that we prefer using dimensionless
parameters constructed with all the dimensional quantities of the
problem instead of fixing some of them to specific values). Therefore,
our critical values $(\epsilon_{c},\eta_{c})=(1,1)$ for the Maxwell
distribution correspond to $(U_{c},\beta_{c})=(3/4,2)$ and our
critical energy $\epsilon_{c}=1/3$ for the water-bag distribution
corresponds to $U_{c}=7/12$, in agreement with \cite{yamaguchi}. Note
furthermore that, within our notations, the stability criterion for
polytropes (Tsallis distributions) takes a very neat form (see
Eq. (156) in \cite{cvb}).

\section{Application}
\label{sec_app}

For illustration, let us apply our results to the numerical
study of \cite{anto}. We consider an initial condition made
of a patch of uniform distribution function $f_{0}=\eta_{0}$ in the
interval ($-\pi\Delta\theta\le \theta\le \pi\Delta\theta$, $-\Delta
v\le v\le \Delta v$) and $f_{0}=0$ outside (water-bag). The density is
$\rho=2\eta_{0}\Delta v$ and the total mass is
\begin{eqnarray}
\label{app1}
M=\eta_{0}4\pi\Delta v\Delta\theta.
\end{eqnarray}
The energy of a distribution which is symmetric with respect to $\theta=0$ is given by
\begin{eqnarray}
\label{app2}
E=\frac{1}{2}\int f v^{2}d\theta dv-\frac{\pi B^{2}}{k},
\end{eqnarray}
where
\begin{eqnarray}
\label{app3}
B=-\frac{k}{2\pi}\int_{0}^{2\pi}\rho(\theta)\cos\theta d\theta,
\end{eqnarray}
is a parameter similar to the magnetization in spin systems
\cite{cvb}. For the water-bag initial condition $f_{0}$, the kinetic
energy is given by
\begin{eqnarray}
\label{app4}
K_{0}=\frac{2}{3}\pi\eta_{0}\Delta\theta (\Delta v)^{3},
\end{eqnarray}
and the magnetization by
\begin{eqnarray}
\label{app5}
B_{0}=-\frac{2k}{\pi}\eta_{0}\Delta v \sin(\pi\Delta\theta).
\end{eqnarray}
It is convenient to introduce the dimensionless parameters
\begin{eqnarray}
\label{app6}
x=\pi\Delta\theta,  \quad y=\Delta v \left (\frac{8}{\pi k M}\right )^{1/2}, \quad b=-\frac{2\pi B_{0}}{kM}.
\end{eqnarray}
Then, the dimensionless initial magnetization can be written
\begin{eqnarray}
\label{app7}
b=\frac{\sin x}{x},
\end{eqnarray}
the dimensionless energy
\begin{eqnarray}
\label{app8}
\epsilon=\frac{\pi^{2}}{6}y^{2}-2b^{2},
\end{eqnarray}
and the degeneracy parameter
\begin{eqnarray}
\label{app9}
\mu=\frac{1}{xy}.
\end{eqnarray}
For a given energy $\epsilon$, the previous relations relate the
initial magnetization $b$ to the degeneracy parameter $\mu$.  We note
this function $b_{\epsilon}(\mu)$. It is represented in
Fig. \ref{application} for a particular value of the energy (see
below). According to the stability diagram of Fig. \ref{muepsilon},
the homogeneous Lynden-Bell distribution is stable for any value of
the degeneracy parameter if $\epsilon\ge 1$. On the other hand, it is
always unstable (i.e., it is not a maximum entropy state) if
$\epsilon<1/3$. Finally, if $1/3\le \epsilon\le 1$, the homogeneous
phase is stable only for $\mu_{min}(\epsilon)\le
\mu\le\mu_{crit}(\epsilon)$ where
$\mu_{min}(\epsilon)=1/\sqrt{6\epsilon}$. Using the curve
$b_{\epsilon}(\mu)$, we can express this stability criterion in terms
of the initial magnetization. We first note that $\Delta\theta\in
\lbrack 0,1\rbrack$ so that $0\le x\le \pi$. On the other hand, for
$x=\pi$, we find that $b=0$ and $\epsilon=(\pi y)^{2}/6$ leading to
$\mu=1/\sqrt{6\epsilon}=\mu_{min}(\epsilon)$. Therefore, if $1/3\le
\epsilon\le 1$, the homogeneous phase is stable only for $0\le b\le
b_{crit}(\epsilon)$ where
$b_{crit}(\epsilon)=b_{\epsilon}(\mu_{crit})$. Above the critical
magnetization $b_{crit}(\epsilon)$, the homogeneous Lynden-Bell
distribution is not a maximum entropy state (it is a saddle point).

\begin{figure}
\centering
\includegraphics[width=8cm]{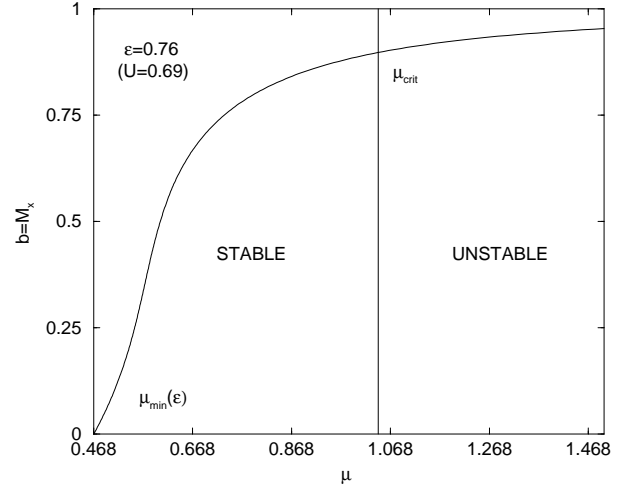}
\caption{Initial magnetization $b=M_{x}$ as a function of the degeneracy parameter $\mu$ for a given value of the energy. There exists a critical magnetization, corresponding to $\mu_{crit}(\epsilon)$, above which the homogeneous Lynden-Bell distribution is unstable.   }
\label{application}
\end{figure}

Using the notations of \cite{anto}, we have
\begin{eqnarray}
\label{app10}
\epsilon=4\left (U-\frac{1}{2}\right ), \qquad b=M_{x}.
\end{eqnarray}
As in \cite{anto}, we fix an energy $U=0.69$. In our notations, it corresponds
to $\epsilon=0.76$. From Fig. \ref{muepsilon}, we find that the
critical degeneracy parameter corresponding to this energy is
$\mu_{crit}=1.043...$. The homogeneous phase is stable for $0.468\le
\mu\le 1.043$. Using the relation between $\mu$ and the initial magnetization
$b=M_{x}$ for $U=0.69$ represented in Fig. \ref{application}, we
conclude that the homogeneous Lynden-Bell distribution is stable
(maximum entropy state) for $M_{x}\le 0.897$. Above this critical
value, it is not a maximum entropy state so it should not result from
a complete violent relaxation: (i) because it is not the most mixed
state (ii) because this distribution is dynamically unstable with
respect to the Vlasov equation. The existence of a critical
magnetization is natural.  Indeed, for $M_{x}\rightarrow 1$, the
degeneracy parameter $\mu\rightarrow +\infty$. Since the Lynden-Bell
distribution becomes ``non-degenerate'' in this limit, the stability
criterion is $U>U_{c}=3/4$ as for the Maxwell-Boltzmann
distribution. Since $U=0.69<U_{c}$ in the simulations, there should
exist a critical magnetization $M_{x,crit}$ above which the system
becomes unstable.  In fact, we already know that the homogeneous
Maxwell-Boltzmann distribution is not always a maximum entropy
state. Since this is a particular limit of the Lynden-Bell
distribution, it is expected that the homogeneous Lynden-Bell
distribution itself is not always a maximum entropy state.

If we consider the inhomogeneous Lynden-Bell distribution
(\ref{vr9}) and assume, without loss of generality, that it is symmetric
with respect to $\theta=0$ (chosen as the origin), the mass, the
magnetization and the energy can be expressed as
\begin{eqnarray}
\label{app11}
M=\eta_{0} \left (\frac{2}{\beta}\right )^{1/2}\int_{0}^{2\pi}I_{-1/2}(\lambda e^{\beta B\cos\theta})\, d\theta,
\end{eqnarray} 
\begin{eqnarray}
\label{app12}
B=-\frac{kM}{2\pi}\frac{\int_{0}^{2\pi}I_{-1/2}(\lambda e^{\beta B\cos\theta})\cos\theta \, d\theta}{\int_{0}^{2\pi}I_{-1/2}(\lambda e^{\beta B\cos\theta}) \, d\theta},
\end{eqnarray}
\begin{eqnarray}
\label{app13}
E=\frac{1}{2}\eta_{0} \left (\frac{2}{\beta}\right )^{3/2}\int_{0}^{2\pi}I_{1/2}(\lambda e^{\beta B\cos\theta})\, d\theta-\frac{\pi B^{2}}{k},
\end{eqnarray} 
where we have used $\Phi=B\cos\theta$.  Equations
(\ref{app11})-(\ref{app13}) determine the equilibrium magnetization
$B$ as a function of the temperature $T$ or energy $E$ for a given value of the
degeneracy parameter $\mu$. The homogeneous state $B=0$ is always a
solution of these equations and we recover
Eqs. (\ref{cc4})-(\ref{cc9}). On the other hand, in the dilute limit
$\lambda\rightarrow +\infty$, using the expansion (\ref{d4}),
Eq. (\ref{app12}) reduces to the implicit equation (25) of
\cite{cvb} for the Maxwell-Boltzmann distribution. To determine the
critical temperature at which the inhomogeneous solution appears, we
expand the equations (\ref{app11})-(\ref{app12}) for $B\rightarrow 0$
as discussed in
\cite{cvb} at a more general level. This yields
\begin{eqnarray}
\label{app14}
B=-\frac{kM}{4\pi}\frac{\lambda I_{-1/2}'(\lambda)}{I_{-1/2}(\lambda)}\beta B\left  \lbrack 1+J(\lambda) (\beta B)^{2}+...\right\rbrack,
\end{eqnarray}
where 
\begin{eqnarray}
\label{app14b}
J(\lambda)=\frac{1}{8}+\frac{3\lambda}{8}\frac{I''}{I'}+\frac{\lambda^{2}}{8}\frac{I'''}{I'}-\frac{\lambda}{4}\frac{I'}{I}-\frac{\lambda^{2}}{4}\frac{I''}{I}, 
\end{eqnarray}
and we have noted $I$ for $I_{-1/2}(\lambda)$. It can be shown that
$J(\lambda)$ is always negative with $J(0)=0$ (Fermi) and
$J(+\infty)=-1/8$ (Maxwell). Therefore, Eq. (\ref{app14}) will have a solution $B\neq 0$ 
if, and only, if 
\begin{eqnarray}
\label{app15}
-\frac{kM}{4\pi}\frac{\lambda I_{-1/2}'(\lambda)}{I_{-1/2}(\lambda)}\beta>1.
\end{eqnarray}
This is precisely the condition (\ref{vs5}) giving the point at which
the homogeneous phase becomes unstable (i.e. becomes a saddle point of
entropy).  Therefore, in continuity with the Maxwell-Boltzmann
distribution (see, e.g., \cite{cvb}), the inhomogeneous Lynden-Bell
phase appears precisely when the homogeneous Lynden-Bell phase becomes
unstable. This result is in fact quite general as discussed in
Sec. 4.3 of \cite{cvb}.  For the Maxwell-Boltzmann distribution, it
has been shown in \cite{cvb} by an explicit calculation (solving an eigenvalue
equation associated with the second order variations of entropy) or by using the Poincar\'e argument for linear series of equilibria that
the inhomogeneous distribution, when it exists, is always a maximum
entropy state. By continuity, we expect in the present case that the
inhomogeneous Lynden-Bell distribution is a maximum entropy state for
the functional (\ref{vr4}). Therefore, in the situation considered in \cite{anto}, the homogeneous Lynden-Bell distribution is a maximum entropy state for $M_{x}<0.897$ and it becomes an unstable saddle point above this critical magnetization. For $M_{x}>0.897$ the maximum entropy state is an inhomogeneous Lynden-Bell distribution. 

The existence of a critical initial magnetization is interesting
because there has been recent claims, based on numerical simulations,
that the dynamics should be different for small ($M0$) and large
($M1$) magnetizations \cite{plr}. Our theoretical study gives further
support to that claim. For $M_{x}<0.897$, our study based on the
statistical mechanics of violent relaxation predicts that the system
should tend to an homogeneous Lynden-Bell distribution with gaussian
tails. This is confirmed by the numerical work of Antoniazzi et
al. \cite{anto} who show simulations up to $M_{x}=0.7$. In that case,
all the results can be explained by standard statistical mechanics and
kinetic theory \cite{bouchet,cvb,bd,pa1,lemou,kin}. However, for
$M_{x}>0.897$, the situation is different. The statistical mechanics
of violent relaxation predicts that the system should tend to an
inhomogeneous Lynden-Bell distribution (most mixed).  The occurence of
a change of regime (dynamical phase transition) seems to be consistent
with numerical results of Pluchino et al. \cite{plr} who observe the
appearance of {\it structures} in the $\mu$-space for large initial
magnetization (but not for small). The fact that $M_{x,crit}=0.897$ is
close to one is also in qualitative agreement with recent reports of
Pluchino
\& Rapisarda \cite{prnew}. When the system becomes spatially inhomogeneous
or involves phase space structures, the statistical mechanics and
kinetic theory become complicated and may lead to anomalies as
mentioned in \cite{curious}. For $M_{x}>0.897$, since the dynamics
becomes more complex, it is possible that the system will not mix well
during violent relaxation so that the inhomogeneous Lynden-Bell
distribution (most mixed state) will {\it not} be achieved in
practice. This is in fact what is observed. Indeed, for $M_{x}=1$, the
degeneracy parameter $\mu\rightarrow +\infty$ so that the Lynden-Bell
distribution coincides with the Maxwell-Boltzmann distribution (non
degenerate limit). Now, for $M_{x}=1$ the QSS is {\it not} a gaussian
as shown in
\cite{latora}. Thus, we expect that for $M_{x}>0.897$ the system will
be trapped in an {\it incompletely mixed} state (which may be
spatially homogeneous or weakly inhomogeneous). This is a stable
stationary solution of the Vlasov equation but different from the
Lynden-Bell distribution due to incomplete relaxation.  Numerical simulations \cite{latora} show that, in certain cases, this state can be fitted by a Tsallis distribution. The Tsallis distributions form a {\it particular}
one-parameter family of stationary solutions of the Vlasov equation
(indexed by $q$) known as stellar polytropes in astrophysics
\cite{pre,cstsallis}. However, there is absolutely {\it no fundamental
reason} why the Tsallis distributions should be selected in a {\it
universal manner} as a result of incomplete violent relaxation
\cite{pre,next05}. Other fits can work as well or even better (see,
e.g., \cite{av}) depending on the situation. In fact, {\it any}
distribution function of the form $f=f(\epsilon)$ where
$\epsilon=v^{2}/2+\Phi$ is the individual energy ($\Phi=0$ for a
homogeneous system) is a steady state of the Vlasov equation.
Furthermore, if $f(\epsilon)$ maximizes an H-function $H=-\int
C(f)d\theta dv$ (where $C$ is convex) at fixed mass $M$ and energy
$E$, then it is nonlinearly dynamically stable with respect to the
Vlasov equation \cite{cvb}. Tsallis distributions associated with
$C(f)=(1/(q-1))(f^q-f)$ are a special case of distribution functions
enjoying these properties but infinitely many other distributions can
be considered as well. The {\it non-universality} of the distribution
function of the QSS, depending on the dynamics, is further discussed
in the Conclusion and in
\cite{next05}. Although the Tsallis distributions have not a fundamental justification in the context of incomplete violent relaxation, an interest 
of these distributions is their {\it mathematical simplicity} and this
is why it is convenient to try this fit first. Such fits have indeed
been considered in \cite{latora} and the best fit is obtained for
$q_{*}=7$
\footnote{The nonlinear dynamical stability of the homogeneous Tsallis
distributions (polytropes) can be studied using the results of
\cite{cvb} (independently, an equivalent calculation has been made in
parallel in \cite{av}). The $q_{*}$ of \cite{latora} is related to our
$q$ in \cite{cvb} by $1-q_{*}=q-1$ hence $q=2-q_{*}=-5$. This
corresponds to a polytropic index $n=1/2+1/(q-1)=1/3$ or
$\gamma=1+1/n=4$. Our study in \cite{cvb} shows that such DF are
nonlinearly dynamically stable with respect to the Vlasov equation
(they maximize the Tsallis $H$-function
\cite{cstsallis} at fixed mass and energy) if
$\epsilon>\epsilon_{crit}=1/\gamma=1/4$ hence $U>9/16=0.5625$ which is
fulfilled for $U=0.69$. By contrast, the homogeneous Maxwell-Boltzmann
distribution ($q=1$) is stable for $U>3/4$ which is not
fulfilled. Therefore, the homogenenous Tsallis distribution with
$q_{*}=7$ is stable while the homogeneous Lynden-Bell distribution
(equivalent to the Maxwell-Boltzmann distribution in the dilute limit
$M_{x}=1$, $\mu\rightarrow +\infty$) is unstable. Note, finally, that
the critical energy $\epsilon_{crit}=1/\gamma$ obtained in \cite{cvb}
is equivalent to the critical energy
$U_{crit}=3/4+(q_{*}-1)/(2(5-3q_{*}))$ obtained in \cite{av} but
expressed in a simpler form.}. It gives a reasonable, but not perfect,
agreement with observations. As said previously, there is no
fundamental reason why the QSS should be precisely described by a
Tsallis distribution, so the deviation between observation and fit
should not cause surprise.  In our opinion, the detailed structure of
the QSS is unpredictable in case of incomplete violent relaxation (see
Conclusion and \cite{next05}). This unpredictability is already
present in the Tsallis $q$ parameter which is a free parameter which
has to be adjusted to the situation in order to fit the results at
best. More generally, we argue that the pdf of the QSS should not
always be a Tsallis distribution, even in case of incomplete
relaxation (i.e. when the Lynden-Bell prediction fails). More general
distributions could arise
\cite{pre,cstsallis,next05}. However, the Tsallis formalism is nice
because it provides {\it simple} analytical expressions of
non-standard distributions (with one single parameter $q$) that can be
handled easily (hence its popularity!). Furthermore, an effective
kinetic theory, based on generalized Fokker-Planck equations, can be
developed in consistency with these distributions
\cite{pp,bukman}. Although more general kinetic theories can be developed
as well for other equilibrium distributions associated with other
forms of ``generalized entropies'' \cite{pre,kingen}, the
$q$-Fokker-Planck equations associated with the Tsallis entropy
provide a good basis for practical studies (due, again, to their {\it
simplicity}) and they may be representative of more general
situations. As suggested in
\cite{kingen,super,pa1}, these {\it effective} kinetic theories could
be useful precisely when standard kinetic theories break down or
become complicated \cite{curious} due to the emergence of structures
in $\mu$-space, non-ergodicity, memory effects, finite $N$ effects
etc. Of course, when standard kinetic theory applies
\cite{bouchet,cvb,bd,pa1,lemou,kin}, they are not necessary. A plausible scenario is that standard statistical mechanics (based, however, on Lynden-Bell's approach) applies for  $M_{x}<0.897$ and that anomalies appear for $M_{x}>0.897$ because the evolution is non-ergodic and involves 
phase-space structures, memory effects etc. It is in this regime that
Tsallis effective thermodynamics (or more general approaches
\cite{pre,kingen}) could be applied. Therefore, our study tends to
reconcile two groups of researchers by suggesting that they study two
different dynamics, below \cite{anto} and above
\cite{plr} the critical initial magnetization $M_{x}=0.897$ (for
$U=0.69$). This is consistent with our general claim
\cite{pre,kingen,cstsallis,super,pa1,next05} that the Tsallis 
formalism can be useful in some situations even if it provides, in our
opinion, only an {\it effective} description of complex systems.
Things are more complex than often said and they deserve a detailed
and careful discussion.

\section{Conclusion}
\label{sec_conclusion}

In this paper, we have investigated the stability of the spatially
homogeneous Lynden-Bell distribution (\ref{cc1}) by determining
whether it corresponds to a maximum of the functional (\ref{vr4}) at
fixed mass and energy. This maximization problem provides a condition
of thermodynamical stability for the process of violent relaxation (in
which case $S$ is interpreted as an entropy) as well as a condition of
nonlinear dynamical stability with respect to the Vlasov equation (in
which case $S$ is interpreted as a generalized $H$-function in the
sense of \cite{tremaine}). The thermodynamical stability condition
ensures that the system is the most mixed state with respect to
microscopic (fine-grained) perturbations. The nonlinear dynamical
stability condition ensures that the coarse-grained distribution is
robust for the collisionless dynamics against macroscopic
perturbations. It is particularly interesting to note that the
conditions of thermodynamical (in Lynden-Bell's sense) and nonlinear
dynamical stability coincide.  We have obtained the expression of the
critical energy and critical temperature above which the homogeneous
phase is stable as a function of the degeneracy parameter. The known
stability criteria corresponding to the Maxwell-Boltzmann distribution
and the water-bag distribution are recovered as particular limits of
our study. Furthermore, a critical point $\mu_{*}=0.68786...$ has been
found below which the homogeneous phase is always stable, whatever the
value of energy and temperature.  When the homogeneous Lynden-Bell
distribution is unstable (saddle point of entropy) it cannot be
achieved as a result of violent relaxation. In that case, the maximum
entropy state is an inhomogeneous Lynden-Bell distribution. For a
given value of energy, the transition should occur for a sufficiently
large value of the initial magnetization ($M_{x}>0.897$ for
$U=0.69$). We have suggested that the relaxation becomes incomplete
for $M_{x}>0.897$ so that the Lynden-Bell prediction fails (this
should be checked numerically but this seems to be the case at least
for $M_{x}=1$ \cite{latora}). In that case, other distributions, that
are stable stationary solutions of the Vlasov equation, can emerge. In
some situations, the QSS can be fitted by the Tsallis distribution
(polytrope). This provide a simple characterisation of the QSS but
this fit is not expected to be universal or fundamental. The fact that
the Lynden-Bell prediction breaks down is the mark of a lack of
ergodicity and a lack of efficient mixing. Dynamical anomalies then
appear and the Tsallis formalism could be used in that context (as an
{\it effective} description) as considered in \cite{latora}. Thus,
having evidenced a new dynamical phase transition, our study may
reconcile two groups of researchers by showing that they describe in
fact different dynamical processes: for $M_{x}<0.897$ the dynamics is
ergodic and ordinary statistical mechanics (Lynden-Bell \cite{lb})
applies and for $M_{x}>0.897$ the dynamics is non-ergodic and {\it
effective} generalized thermodynamics, such as Tsallis \cite{tsallis}
approach (or more general
\cite{pre,kingen} approaches) could be tried.

We would like to conclude this study by emphasizing the limitations of
the Lynden-Bell statistical theory at a general level. First of all,
the distribution function (\ref{vr9}) is only valid when the initial
condition takes two values $0$ and $\eta_{0}$. Therefore, it is not
expected to apply to any initial condition. For more complex initial
conditions, the Lynden-Bell prediction is a superposition of
Fermi-Dirac distributions for all the phase levels $\eta$. Therefore,
depending on the initial conditions, the Lynden-Bell distribution can
take a wide diversity of forms given by the general formula
(\ref{vr12}); this is similar to a sort of superstatistics
\cite{super} where $\chi(\eta)$ is fixed by the initial condition. The Fermi-Dirac distribution (\ref{vr9}) is just a
particular case of this general formula for two levels.  It is indeed
important to stress that the prediction of Lynden-Bell depends on the
{\it details} of the initial condition, not only on the robust
conserved quantities $E$ and $M$. This is at variance with usual
statistical mechanics where only the robust contraints (energy,
mass,...) matter. This is due to the existence of Casimir contraints
in the Vlasov dynamics that act as {\it hidden constraints}
\cite{super}. When we consider realistic initial conditions, we enter
into complications because: (1) we need to discretize the initial
condition into several levels and then relate the Lagrange multipliers
$\chi(\eta)$ to the hypervolume $\gamma(\eta)$ of each level
\cite{csr,super}. This makes the application of the Lynden-Bell theory
technically complicated and heavy because it involves a lot of control
parameters \cite{brands}. (2) we do not always know at which scale we
must discretize the initial condition and different discretizations
may lead to different results as discussed in \cite{jfm1} (p. 284) and
in \cite{arad}. (3) in addition, there is a debate to decide whether
all the Casimirs are conserved (microscopically) or if certain are
altered by non-ideal effects during the dynamics so that a {\it prior}
distribution should be introduced instead (this remark applies
particularly to realistic systems such as 2D turbulence
\cite{ellis,pre,physicaD,kupka,leprovost}). The simple two-levels situation
considered in Antoniazzi et al. \cite{anto} is not subject to such
difficulties and criticisms.

On the other hand, the approach of Lynden-Bell assumes that the system
mixes well so that the hypothesis of {\it ergodicity} \footnote{It
should be clear that, in this paper, we are talking about the
ergodicity with respect to the collisionless mixing in relation with
the process of violent relaxation, not the ergodicity with respect to
the collisional evolution. Collisional relaxation (due to
granularities and finite $N$ effects) is not considered here because
it occurs on very long timescales for $N\gg 1$ and does not account
for the structure of the QSS \cite{cvb}.} which sustains his statistical theory
(maximization of the entropy) is fulfilled. Again, this is not
expected to be general. Several cases of {\it incomplete} violent
relaxation have been identified in stellar dynamics and 2D turbulence
(see some references in
\cite{next05}). In such cases, the QSS is not exactly described by the
Lynden-Bell distribution but it is nevertheless a stable stationary
solution of the Vlasov equation. Therefore, distributions different
from (\ref{vr9}) can emerge in case of incomplete violent
relaxation. This is the case for example in the plasma experiment of
Huang \& Driscoll \cite{hd} in 2D turbulence where it has been
observed that the QSS is not perfectly well-described by the Lynden-Bell
theory. In particular, the density drops to zero at a finite
distance instead of decaying smoothly. Boghosian \cite{boghosian} has
interpreted this result in terms of Tsallis non-extensive
thermodynamics. Alternatively, the deviation from the Lynden-Bell
prediction has been interpreted in
\cite{brands} as a result of an incomplete violent relaxation (non-ergodicity)
and a lack of mixing in the core and the halo of the ``vortex''. In
this interpretation, the QSS is viewed as a particular stable
stationary solution of the 2D Euler equation which is not the most
mixed state. As we understand, the proposal of Tsallis is to introduce
an entropy which can take into account {\it non-ergodic} behaviours
\footnote{In this respect, we should stress that the proper form of
Tsallis entropy for the process of violent relaxation is the one given
in \cite{next05}, expressed in terms of $\rho(\theta,v,\eta)$. In the
two-levels approximation, it reduces to a $q$-Fermi-Dirac type entropy
\cite{brands}. Within this interpretation, the index $q$ would 
be a measure of mixing. For $q=1$ (efficient mixing), we recover the
Lynden-Bell theory (see
\cite{next05}). }. Indeed, generalized entropies make sense only when
the system is non ergodic. If the system mixes well, as in the
situations considered in \cite{anto}, the Lynden-Bell theory is the
relevant one \cite{next05}. Unfortunately, we do not
know {\it a priori} whether the system will mix well or not; this
depends on the dynamics and on the route to equilibrium
\cite{next05}. We only know {\it a posteriori} if
the Lynden-Bell prediction has worked or failed (like in, e.g.,
\cite{hd}). The possibility that non-ergodic behaviours could be
described by a generalized form of entropy is an attractive
idea. However, we do {\it not} believe that complicated non-ergodic
behaviours associated with the process of incomplete violent
relaxation can be encapsulated in a simple functional such as the
$q$-entropy proposed by Tsallis. When the system does not mix well we
can have a wide variety of QSS. This is because the Vlasov (or 2D
Euler) equation admits an infinite number of steady states and the
system can be trapped in one of them. A kinetic theory of violent
relaxation, as initiated in \cite{csr,quasi}, is then necessary to
account for incomplete relaxation (see \cite{next05}).

In conclusion, for these two reasons: (1) dependence on the detailed
structure of the initial conditions (through the Casimirs constraints)
(2) incomplete violent relaxation (non ergodicity), the QSS is {\it
not} expected to be described by a ``universal'' distribution such as
(\ref{cc1}), or even (\ref{vr12}).  However, as claimed long ago in
\cite{brands}, the Lynden-Bell theory is the only one to make a {\it
prediction} of the QSS without ad hoc parameter. It is usually
observed that the Lynden-Bell prediction provides a {\it fair}
description of the QSS in many cases even if there can exist
discrepencies that are more or less pronounced due to incomplete
relaxation. Thus, the Lynden-Bell distribution gives the tendency to
which state the system {\it should} tend as a result of
mixing. However, during the route to equilibrium, mixing may not be
sufficient and the system can be frozen in a stationary solution of
the Vlasov equation which is not the most mixed state. This process of
incomplete relaxation is beautifully described by Binney \& Tremaine
\cite{bt}, pp. 266-267, using an analogy with the structure of the Mississippi
river. On the other hand, Tsallis distributions form just a
particular one-parameter family of stationary solutions of the Vlasov
equation (analogous to stellar polytropes) that can sometimes be used
as convenient {\it fits}, in case of incomplete violent relaxation, due to
their simple mathematical expression. However, these fits should not
work in a universal manner and other fits can work as well, or even
better.

There are still debates and controversies about the fact that the
collisionless evolution of the HMF model is adequately described by
the Vlasov equation.  Indeed, the authors of \cite{latora,plr,rp}
describe the QSS in terms of Tsallis generalized thermodynamics based
on the $N$-body system, not in terms of the Lynden-Bell thermodynamics
based on the Vlasov equation. The Vlasov description is classical in
plasma physics and astrophysics (as well as in point vortex dynamics
and 2D turbulence where it has the form of the Euler equation
\cite{houches}) to describe the evolution of the system during the
regime where ``collisions'' are negligible. In that case, there are no
correlations between the particles and the $N$-body distribution
function is a product of $N$ one-body distribution functions. Using
this property to close the BBGKY hierarchy steming from the Liouville
equation, one obtains the Vlasov equation
\cite{pa1}. For a system with weak long-range interactions, it can be shown
that this mean field description is correct in a proper thermodynamic
limit $N\rightarrow +\infty$ \footnote{As shown in
\cite{bd,cvb,pa1}, the Vlasov equation also describes the collisional
relaxation at order $1/N$. This is because the Lenard-Balescu
collision term cancels out for 1D systems (this property is known for
a long time in plasma physics
\cite{kp}). Collisions therefore manifest themselves on a timescale larger than $N$. For the HMF model, numerical simulations show that the relaxation time scales as $N^{1.7}$ \cite{yamaguchi}.}. A mathematically rigorous derivation of the Vlasov
equation is given by Braun \& Hepp \cite{bh}. However, justifying the
Vlasov equation for the collisionless regime of the HMF model is not
the end of the story. As noted in \cite{next05}, the Vlasov equation
coupled to a long-range force can have a very complicated behaviour
(this is similar to the Euler equation in 2D turbulence
\cite{houches}). Therefore, the anomalies (non-ergodicity, phase space structures,...) observed by
\cite{latora,plr,rp} in their $N$-body simulations would probably persist and be observed by directly solving the Vlasov equation. This would 
describe the $N\rightarrow +\infty$ limit of the model. This
comparative study ($N$-body Hamilton equations versus Vlasov equation)
has not yet been done for the HMF model but it would be an interesting
step to reconcile different approaches. It would also clearly show
whether the ``anomalies'' reported in \cite{latora,plr,rp} are due to
finite $N$ effects or if they persist in the thermodynamic limit
$N\rightarrow +\infty$.

Note finaly that, since the Lynden-Bell distribution (\ref{vr9}) is
similar to the Fermi-Dirac statistics in quantum mechanics, the
results of this paper also describe the (ordinary) statistical
mechanics of a system of $N$ fermions on a ring interacting via a cosine
potential; this could be called the {\it fermionic HMF model}. In that
context, the maximum value of the distribution function is
$\eta_{0}=g/h$ where $h$ is the Planck constant and $g=2s+1$ the spin
multiplicity of the quantum states. In this quantum context, $\eta_{0}$ is
fixed by the Pauli exclusion principle (see the analogous situation
for self-gravitating fermions in \cite{pt}).

\vskip1cm
{\it Acknowledgements} I acknowledge stimulating discussions with
D. Lynden-Bell and C. Tsallis.


\begin{thebibliography}{99}

\bibitem{dauxois}  {\small  Dynamics and Thermodynamics of Systems with Long Range Interactions, edited by T. Dauxois, S. Ruffo, E. Arimondo and M. Wilkens, Lect. Not. in Phys. {\bf 602}, Springer (2002)}

\bibitem{kk}  {\small T. Konishi, K. Kaneko, J. Phys. A {\bf 25}, 6283 (1992)}

\bibitem{inagaki}  {\small S. Inagaki, T. Konishi, Publ. Astron. Soc. Jpn {\bf 45}, 733 (1993)}

\bibitem{pichon}  {\small C. Pichon, Ph.D. thesis, Cambridge (1994)}

\bibitem{ruffo}  {\small M. Antoni, S. Ruffo, PRE {\bf 52}, 2361 (1995)}

\bibitem{latora}  {\small  V. Latora, A. Rapisarda, C. Tsallis, PRE {\bf 64}, 056134 (2001); Physica A  {\bf 305}, 129 (2002) }

\bibitem{choi}  {\small  M.Y. Choi, J. Choi, PRL {\bf 91}, 124101 (2003) }

\bibitem{plr}  {\small A. Pluchino, V. Latora, A. Rapisarda,  Physica D  {\bf 193}, 315 (2004)}

\bibitem{yamaguchi}  {\small Y.Y. Yamaguchi, J. Barr\'e, F. Bouchet, T. Dauxois, S. Ruffo,  Physica A  {\bf 337}, 36 (2004)}

\bibitem{bouchet}  {\small F. Bouchet, PRE  {\bf 70}, 036113 (2004)}

\bibitem{av}  {\small C. Anteneodo, R.O. Vallejos, Physica A  {\bf 344}, 383 (2004)}

\bibitem{cvb}  {\small P.H. Chavanis, J. Vatteville, F. Bouchet, Eur. Phys. J. B {\bf 46}, 61 (2005)}

\bibitem{kaneko}  {\small H. Morita, K. Kaneko, PRL {\bf 94}, 087203 (2005)}

\bibitem{rf}  {\small T.M. Rocha Filho, A. Figueiredo, M.A. Amato, PRL {\bf 95}, 190601 (2005)}

\bibitem{rp}  {\small A. Rapisarda, A. Pluchino, Europhysics News  {\bf 36}, 202 (2005)}

\bibitem{bd}  {\small F. Bouchet, T. Dauxois, PRE  {\bf 72}, 045103(R) (2005)}

\bibitem{curious}  {\small P.H. Chavanis,  Eur. Phys. J. B {\bf  52}, 47 (2006)}

\bibitem{campa}  {\small A. Campa, A. Giansanti, D. Mukamel, S. Ruffo, Physica A {\bf 365},  120 (2006)}

\bibitem{pluchino}  {\small A. Pluchino, A. Rapisarda, Physica A {\bf 365}, 184 (2006)}

\bibitem{anteneodo}  {\small L.G. Moyano, C. Anteneodo PRE {\bf 74}, 021118 (2006)}

\bibitem{baldovin}  {\small F. Baldovin, E. Orlandini, PRL {\bf 96},  240602  (2006)}

\bibitem{houches}  {\small  P.H. Chavanis, {\it Statistical Mechanics of Two-dimensional Vortices and Stellar Systems} in \cite{dauxois} 
[cond-mat/0212223]}

\bibitem{pa1}  {\small P.H. Chavanis, Physica A {\bf 361}, 55 (2006); Physica A {\bf 361}, 81 (2006) }

\bibitem{tsallis}  {\small C. Tsallis, J. Stat. Phys.  {\bf 52}, 479 (1988)}

\bibitem{next05}  {\small P.H. Chavanis, Physica A {\bf 365}, 102 (2006)}

\bibitem{anto}  {\small A. Antoniazzi, D. Fanelli, J. Barr\'e, P.H. Chavanis, T. Dauxois, S. Ruffo, [cond-mat/0603813v2] }

\bibitem{lb}  {\small  D. Lynden-Bell, MNRAS  {\bf 136}, 101 (1967)}

\bibitem{csr}  {\small  P.H. Chavanis, J. Sommeria, R. Robert,
ApJ, {\bf 471}, 385 (1996)}

\bibitem{thesis}  {\small  P.H. Chavanis,  {\it Contributions \`a la m\'ecanique statistique des tourbillons bidimensionnels. Analogie avec la relaxation violente des syst\`emes stellaires}, Ph.D. Thesis, ENS Lyon (1996) }

\bibitem{dubrovnik}  {\small  P.H. Chavanis, {\it Statistical Mechanics of Violent Relaxation in Stellar Systems} in: Proceedings of the
     Conference on Multiscale Problems in Science and Technology,
     edited by N. Antonic, C.J. van Duijn, W.  Jager and A. Mikelic
     (Springer, Berlin, 2002) [astro-ph/0212205] }

\bibitem{super}  {\small P.H. Chavanis, Physica A {\bf  359}, 177 (2006) }

\bibitem{csmnras}  {\small P.H. Chavanis, J. Sommeria, MNRAS {\bf 296}, 569 (1998)}

\bibitem{pre}  {\small P.H. Chavanis, PRE {\bf 68}, 036108 (2003)}

\bibitem{bt}  {\small J. Binney, S. Tremaine, {\it Galactic Dynamics} (Princeton Series in Astrophysics, 1987) }

\bibitem{aaa}  {\small P.H. Chavanis, A\&A {\bf 401}, 15 (2003)  } 

\bibitem{tremaine}  {\small S. Tremaine, M. H\'enon, D. Lynden-Bell, MNRAS  {\bf 227}, 543 (1986)}

\bibitem{cstsallis}  {\small P.H. Chavanis, C. Sire, Physica A {\bf 356}, 419 (2005)}

\bibitem{aaantonov}  {\small P.H. Chavanis, A\&A {\bf  451}, 109 (2006) }

\bibitem{miller}  {\small J. Miller, PRL {\bf 65}, 2137 (1990)}

\bibitem{rs}  {\small R. Robert, J. Sommeria, JFM {\bf 229}, 291 (1991)}

\bibitem{aa1}  {\small P.H. Chavanis, A\&A {\bf 381}, 340 (2002)  } 

\bibitem{lemou}  {\small P.H. Chavanis, M. Lemou, PRE {\bf 72}, 061106 (2005)  } 
 
\bibitem{kin}  {\small P.H. Chavanis, Eur. Phys. J. B {\bf 52}, 61 (2006)  } 

\bibitem{prnew}  {\small A. Pluchino, A. Rapisarda, Prog. Theor. Phys. Supp. {\bf 162}, 18  (2006)}

\bibitem{pp}  {\small A.R. Plastino, A. Plastino, Physica A {\bf 222}, 347 (1995)}

\bibitem{bukman}  {\small C. Tsallis, D.J. Bukman, PRE {\bf 54}, R2197 (1996)}

\bibitem{kingen}  {\small P.H. Chavanis, Physica A {\bf 332}, 89 (2004); Physica A {\bf 340}, 57 (2004).}

\bibitem{brands}  {\small H. Brands, P.H. Chavanis, R. Pasmanter, J. Sommeria, Phys. Fluids {\bf 11}, 3465 (1999)}

\bibitem{jfm1}  {\small P.H. Chavanis, J. Sommeria, J. Fluid Mech. {\bf 314}, 267 (1996)}

\bibitem{arad}  {\small I. Arad, D. Lynden-Bell, MNRAS {\bf 361}, 385 (2005)}

\bibitem{ellis}  {\small R. Ellis, K. Haven, B. Turkington, Nonlinearity {\bf 15}, 239 (2002)}

\bibitem{physicaD}  {\small P.H. Chavanis, Physica D {\bf 200}, 257 (2005)}

\bibitem{kupka}  {\small P.H. Chavanis, {\it Statistical Mechanics of 2D Turbulence with a Prior Vorticity Distribution} in: Proceedings of the Workshop on 
Interdisciplinary Aspects of Turbulence at Ringberg Castle, Tegernsee,
Germany (Max-Planck Institut fur Astrophysik, 2005) [physics/0601087]}

\bibitem{leprovost}  {\small N. Leprovost, B. Dubrulle, P.H. Chavanis, PRE  {\bf 73}, 046308 (2006)}

\bibitem{hd}  {\small X.P. Huang, C.F. Driscoll, PRL {\bf 72}, 2187 (1994)}

\bibitem{boghosian}  {\small B.M. Boghosian, PRE {\bf 53}, 4754 (1996)}

\bibitem{quasi}  {\small P.H. Chavanis, MNRAS {\bf 300}, 981 (1998)}

\bibitem{kp}  {\small B.B. Kadomtsev, O.P. Poguste, PRL {\bf 25}, 1155 (1970)}

\bibitem{bh}  {\small W. Braun, K. Hepp, Commun. Math. Phys. {\bf 56}, 101 (1977)}

\bibitem{pt}  {\small P.H. Chavanis, PRE {\bf 65}, 056123 (2002)}
 

\end{thebibliography}
\end{document}